%% file: Nils_Asmussen.tex
\documentclass{PoS}
\usepackage{mathtools,siunitx,booktabs}
\usepackage{tikz,tikz-feynman}
\usepackage{subfig}
\usepackage{graphicx}
\usepackage{bm}
\usepackage[capitalise]{cleveref}

\graphicspath{{PLOTS/}}

\newcommand\mcL{\mathcal{L}}

\newcommand\bmcL{\bar{\mathcal{L}}}

\newcommand{\be}{\begin{equation}}
\newcommand{\ee}{\end{equation}}
\newcommand{\ba}{\begin{eqnarray}}
\newcommand{\ea}{\end{eqnarray}}
\newcommand{\bi}{\begin{itemize}}
\newcommand{\ei}{\end{itemize}}

\newcommand{\<}{\langle} 
\renewcommand{\>}{\rangle}

\FullConference{37th International Symposium on Lattice Field Theory - Lattice2019\\
                16-22 June 2019\\
                Wuhan, China}

\let\oldmaketitle\maketitle
\renewcommand\maketitle{{\bfseries\boldmath\oldmaketitle}}

\title{Developments in the position-space approach to the HLbL contribution to 
the muon $\bm{g-2}$ on the lattice}
\ShortTitle{Position-space approach to the HLbL contribution to the muon $g-2$ on the 
lattice}

\author{%
\speaker{Nils~Asmussen}$^{,a}$,
En-Hung~Chao$^b$,
Antoine~G\'erardin$^c$,
Jeremy~R.~Green$^d$,
Renwick~J.~Hudspith$^b$,
Harvey~B.~Meyer$^{b,e}$ and
Andreas~Nyffeler$^b$\\
\llap{$^a$}School of Physics and Astronomy, University of Southampton\\
Southampton SO17 1BJ, UK\\
\llap{$^b$}PRISMA$^+$ Cluster of Excellence and Inst. f\"ur Kernphysik, Johannes Gutenberg-Universit\"at Mainz\\
D-55099 Mainz, Germany\\
\llap{$^c$}
John von Neumann Institute for Computing, DESY, Platanenallee 6\\
D-15738 Zeuthen, Germany\\
\llap{$^d$}Theoretical Physics Department, CERN\\
CH-1211 Geneva 23, Switzerland\\
\llap{$^e$}Helmholtz Institut Mainz\\
D-55099 Mainz, Germany\\
E-mail:
\email{n.asmussen@soton.ac.uk},
\email{enchao@uni-mainz.de},
\email{antoine.gerardin@desy.de},
\email{jeremy.green@cern.ch},
\email{renwick.james.hudspith@googlemail.com},
\email{meyerh@uni-mainz.de},
\email{nyffeler@uni-mainz.de}%
}

\abstract{ The measurement of the anomalous magnetic moment of the
  muon and its prediction allow for a high-precision test of the
  Standard Model (SM). In this proceedings article we present ongoing
  work combining lattice QCD and continuum QED in order to determine
  an important SM contribution to the magnetic moment, the hadronic
  light-by-light contribution. We compute the quark-connected
  contribution in the Mainz position-space approach and investigate
  the long-distance part of our data using calculations of the $\pi^0$-pole
  and charged pion loop contributions.  }

\begin{document}
\section{Introduction}
One of the most stringent tests of the Standard Model (SM) arises from the 
measurement of the anomalous magnetic moment of the muon $a_\mu$. 
A tension of about three standard deviations persists
between the SM prediction 
for this quantity and its experimentally measured value. The theory and 
experimental uncertainties are comparable and at the sub-ppm level, but new experiments such as the ``E989 
Muon g-2'' at Fermilab and the ``Muon g-2/EDM'' at JPARC expect an improvement in precision 
by about a factor four in the next few years;
see~\cite{Jegerlehner2018} and references therein.  It is necessary to reduce 
the theoretical uncertainty by a comparable amount in order to discern whether 
the current discrepancy between theory and experiment is a sign of Beyond the 
Standard Model physics.

The theoretical uncertainty for $a_\mu$ is currently dominated by hadronic 
contributions, namely the hadronic vacuum polarization (HVP) as well as the hadronic 
light-by-light (HLbL) scattering. It is the latter of these contributions that 
we will focus on in this proceedings article. We will summarize the methodology and 
present some preliminary results for the contribution of the quark-connected diagrams to
$a_\mu^{\text{HLBL}}$ as well as a discussion of finite-volume effects and our 
use of continuum models to describe the long-distance part of our data. To this end, 
continuum computations of the \(\pi^0\)-pole and charged pion loop contributions 
are presented.

\section{Position-space method}
To compute the HLbL contribution, we make use of the Mainz position-space 
method, see also 
references~\cite{Green2016,Asmussen2016,Asmussen2018,Asmussen2018a,Asmussen2019}.  
It divides the problem into a QED part and a QCD part. The QED part is described 
by a kernel function~\(\bmcL\), that is computed in the continuum and infinite 
volume, and the QCD part is given by a four-point function \(i\widehat\Pi\), 
that is to be obtained with the help of Lattice QCD.  The master 
formula that allows one to compute the HLbL contribution to \(a_\mu\) reads
\begin{align}
   a_\mu^{\rm HLbL}
   &= \frac{m e^6}{3} \int d^4y
   \Big[
      \int d^4x
      \underbrace{\bar{\cal L}_{[\rho,\sigma];\mu\nu\lambda}(x,y)}_{\rm QED}\;  
      \underbrace{i\widehat\Pi_{\rho;\mu\nu\lambda\sigma}(x,y)}_{\rm QCD}
   \Big].
   \label{eq:master}
   \\
   i\widehat \Pi_{\rho;\mu\nu\lambda\sigma}( x, y)
   &= \int d^4z\; (-z_\rho)\widetilde\Pi_{\mu\nu\sigma\lambda}(x,y,z), \quad 
 \widetilde\Pi_{\mu\nu\sigma\lambda}(x,y,z)=  \big\<\,j_\mu(x)\,j_\nu(y)\,j_\sigma(z)\, j_\lambda(0)\big\>\,,
\end{align}
where \(m\) is the mass of the muon and the \(j_\mu(x)\) are the quark 
electromagnetic currents.

\begin{figure}[t]
   \includegraphics[width=0.48\textwidth]{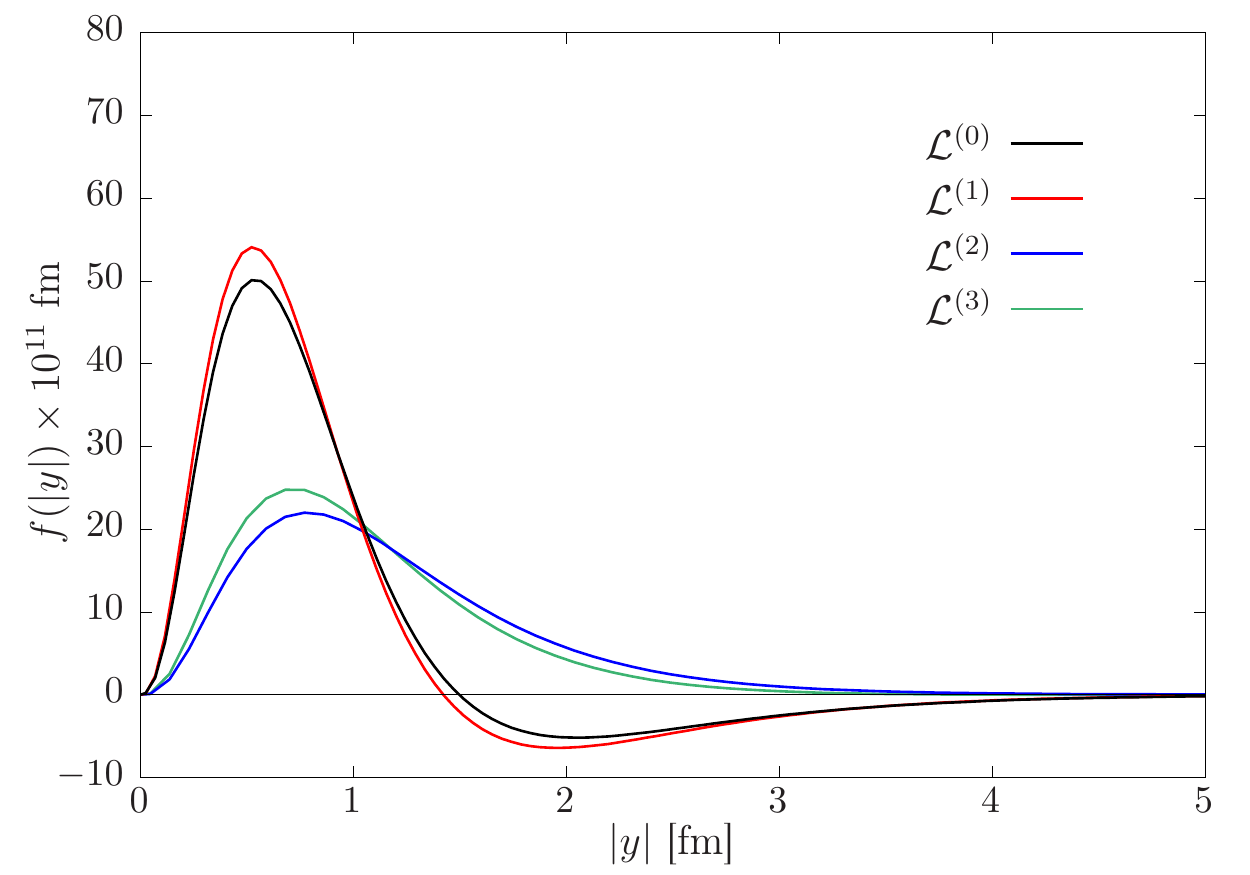}\hfill
   \includegraphics[width=0.48\textwidth]{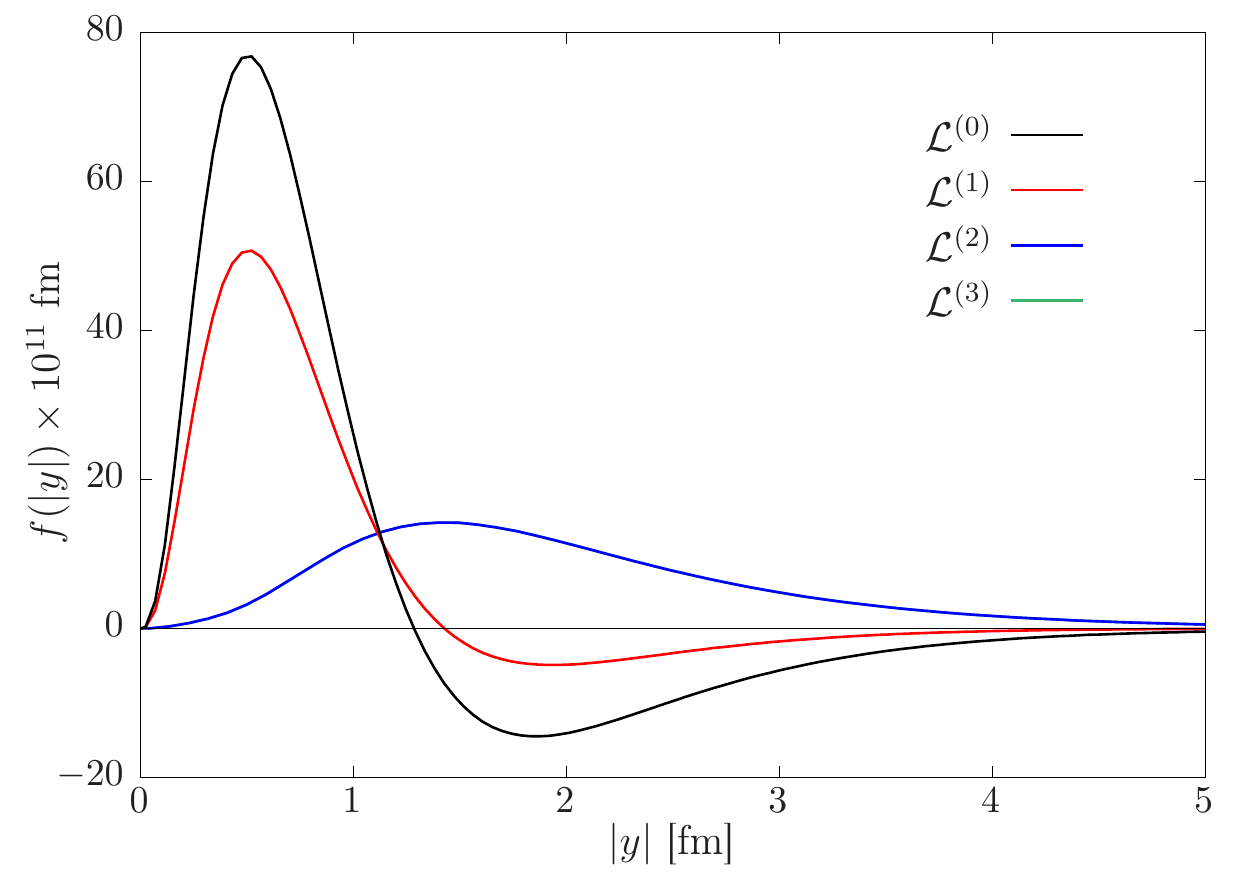}
   \caption{%
      The integrand for the $\pi^0$-pole contribution based on the standard 
      kernel \(\mcL^{(0)}\) and the subtracted kernels \(\mcL^{(1,2,3)}\) at 
      \(m_\pi=\SI{300}{MeV}\), using method 1 (left panel) and method 2 (right 
      panel).  In the right-hand plot, the \(\mcL^{(3)}\) curve is hidden behind 
      the \(\mcL^{(2)}\) curve.  The continuum calculation is performed using 
      the VMD model for the pion transition form factor.%
   }
   \label{fig:subtractions}
\end{figure}

The kernel~\(\bmcL\) is not unique. Other valid kernels can be obtained by 
adding or subtracting terms that vanish after the \(x\) and \(y\) integrations 
in the master formula~\eqref{eq:master}.  Such subtractions were first 
introduced in~\cite{Blum2017}, where it was shown that discretization effects 
can be drastically reduced by choosing kernels that vanish when some of the 
vertices coincide.  Exploiting
\(\int_x i\widehat\Pi(x,y)=\int_y i\widehat\Pi(x,y)=0\,,\)
we have tested the usefulness of the subtracted kernels \(\mcL^{(1-3)}\),
\begin{align}\label{eq:L0}
   \mcL^{(0)}=&\bar{\mcL}(x,y)\,,\quad \text{(standard kernel)}\\
\label{eq:L1}
   \mcL^{(1)}
       =&\bar{\mcL}(x,y)-\frac{1}{2}\bar{\mcL}(x,x)-\frac{1}{2}\bar{\mcL}(y,y)\,,
   \\
\label{eq:L2}
   \mcL^{(2)}
      =&\bar{\mcL}(x,y)-\bar{\mcL}(0,y)-\bar{\mcL}(x,0)\,,
   \\
\label{eq:L3}
   \mcL^{(3)}
      =&\bar{\mcL}(x,y)-\bar{\mcL}(0,y)-\bar{\mcL}(x,x)+\bar{\mcL}(0,x)\,,
\end{align}
that obey the following properties:
\begin{align}
   \mcL^{(0)}(0,0)=0\,,\quad
   \mcL^{(1)}(x,x)=0\,,\quad
   \mcL^{(2)}(0,y)=\mcL^{(2)}(x,0)=0\,,\quad
   \mcL^{(3)}(x,x)=\mcL^{(3)}(0,y)=0\,.
\end{align}
The left panel of~\cref{fig:subtractions} displays the integrands 
$f(|y|)$ of the final integration over~$|y|$ corresponding to the different 
kernels, for the neutral pion pole contribution with a vector-meson-dominance 
(VMD) model of the pion transition form factor.
Compared to the standard kernel, \(\mcL^{(2)}\) and \(\mcL^{(3)}\)
have less pronounced peaks at short distances and approach zero faster
at long distances.  We expect these subtracted kernels to have smaller
lattice artifacts and therefore to be favorable in lattice computations.

\begin{figure}[t]
   \hfill
   \input{Feynman/methods.tex}
   \hfill
   \subfloat[]{
      \begin{tabular}{l}
         \includegraphics[width=0.45\textwidth]{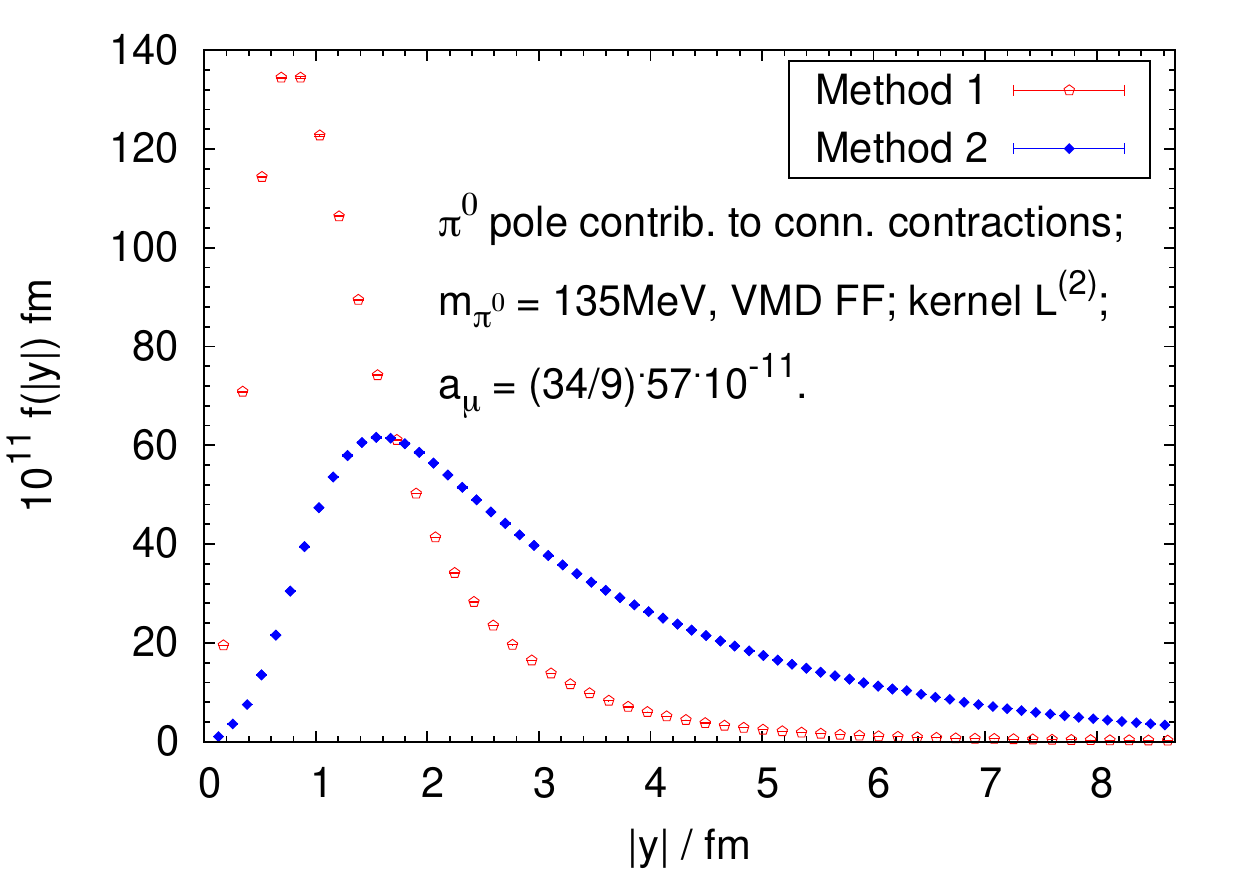}
      \end{tabular}
   }
   \hfill
   \caption{%
      (a) Contractions needed to compute \(g-2\). Upper row: The three connected 
      Wick contractions needed in method 1.  Bottom row: In method 2 the 
      different contraction types are implemented in the QED part of the 
      diagram. (b) Comparison of the integrands for method 1 and method 2 for 
      the neutral pion pole contribution with a transition form factor given by the VMD 
      model.
   }
   \label{fig:methods}
\end{figure}

The quark-connected part of the four-point function \(i\widehat\Pi\),
 involves three different contractions. Computing all three of them and applying Eq.\ (\ref{eq:master}) 
amounts to what we call `method 1'. In a lattice 
implementation of this method, for \(N\) evaluations of the \(y\) integrand, 
\(1+N\) propagators and \(6(1+N)\) sequential propagators are needed.  
If $\Pi^{(1)}_{\mu\nu\sigma\lambda}(x,y,z)\equiv -2{\rm ReTr}\{ S(0,x)  \gamma_\mu S(x,y)  \gamma_\nu S(y,z)  \gamma_\sigma S(z,0) \gamma_\lambda \}$, where the \(S(x,y)\) are propagators, 
represents one of the Wick contractions of the quark-connected part $\widetilde\Pi^c_{\mu\nu\sigma\lambda}(x,y,z)$,
we can write (for any given background gauge field)
\begin{align}
\widetilde\Pi^c_{\mu\nu\sigma\lambda}(x,y,z) = 
   \Pi^{(1)}_{\mu\nu\sigma\lambda}(x,y,z)+\Pi^{(1)}_{\nu\mu\sigma\lambda}(y,x,z)+\Pi^{(1)}_{\nu\sigma\mu\lambda}(y,z,x).
\end{align}
Note that $\partial_\mu(x) \widetilde \Pi^c_{\mu\nu\sigma\lambda}(x,y,z) = 0$ for all $x$.
The computation can be arranged in a different way, such that only the contraction $\Pi^{(1)}$ is computed
and the others are implemented by permuting the way that the photons are attached to the 
vertices of the four-point function. We call this method 2, which reads
\begin{align}
   a_\mu^{\text{HLbL,c}} = \frac{me^6}{3}
   \int_{y,x,z}\Big(&
      [\bmcL_{[\rho,\sigma];\mu\nu\lambda}(x,y)
      +\bmcL_{[\rho,\sigma];\nu\mu\lambda}(y,x)
      -\bmcL_{[\rho,\sigma];\lambda\nu\mu}(x,x-y)]
      (-z_\rho)\,\Pi^{(1)}_{\mu\nu\sigma\lambda}(x,y,z)
\nonumber\\ &
      +\bmcL_{[\rho,\sigma];\lambda\nu\mu}(x,x-y)\,(-x_\rho) \Pi^{(1)}_{\mu\nu\sigma\lambda}(x,y,z)
   \Big).
\end{align}
A diagrammatic representation and the integrands for both methods are shown 
in~\cref{fig:methods,fig:subtractions}. While method 2 requires the calculation 
of far fewer propagators, its integrand receives contributions from the exchange 
of resonances odd under charge conjugation, which cancel out upon fully 
integrating over \(x,y,z\).

\section{Lattice results}
The results described in this section are obtained with kernel
\(\mcL^{(2)}\), which we expect to reduce lattice artifacts.  For our
lattice calculations, which are based on the ensembles listed
in~\cref{tab:ensembles}, we use method 2 to reduce the number of
required inversions of the Dirac operator.  From Fig.~3, we observe
that we achieve good statistical precision for small $|y|$, but at
larger distances the signal degrades rapidly.

\begin{figure}[h!]
   \subfloat[\label{fig:discretization}]{\includegraphics[width=0.48\textwidth]{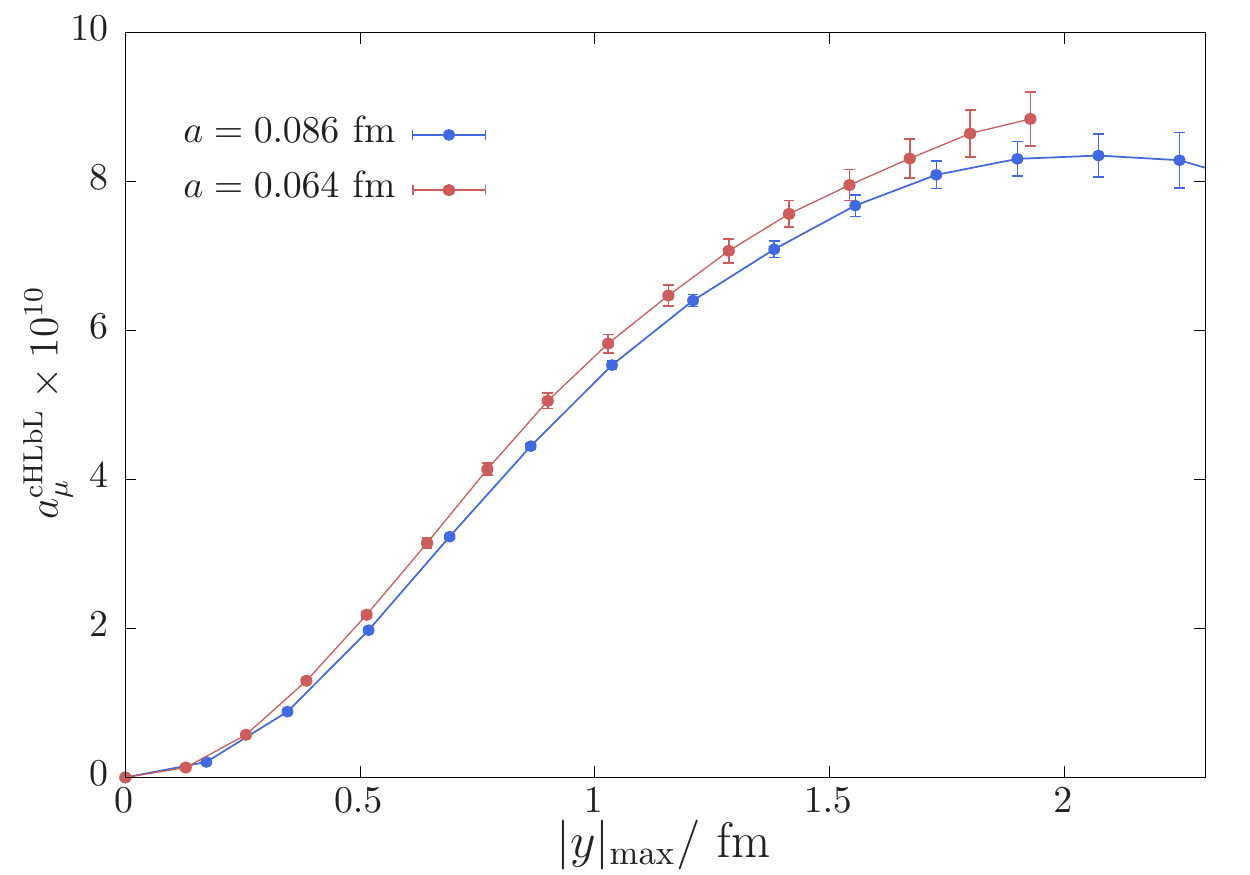}}
   \hfill
   \subfloat[\label{fig:finitesize}]{\includegraphics[width=0.48\textwidth]{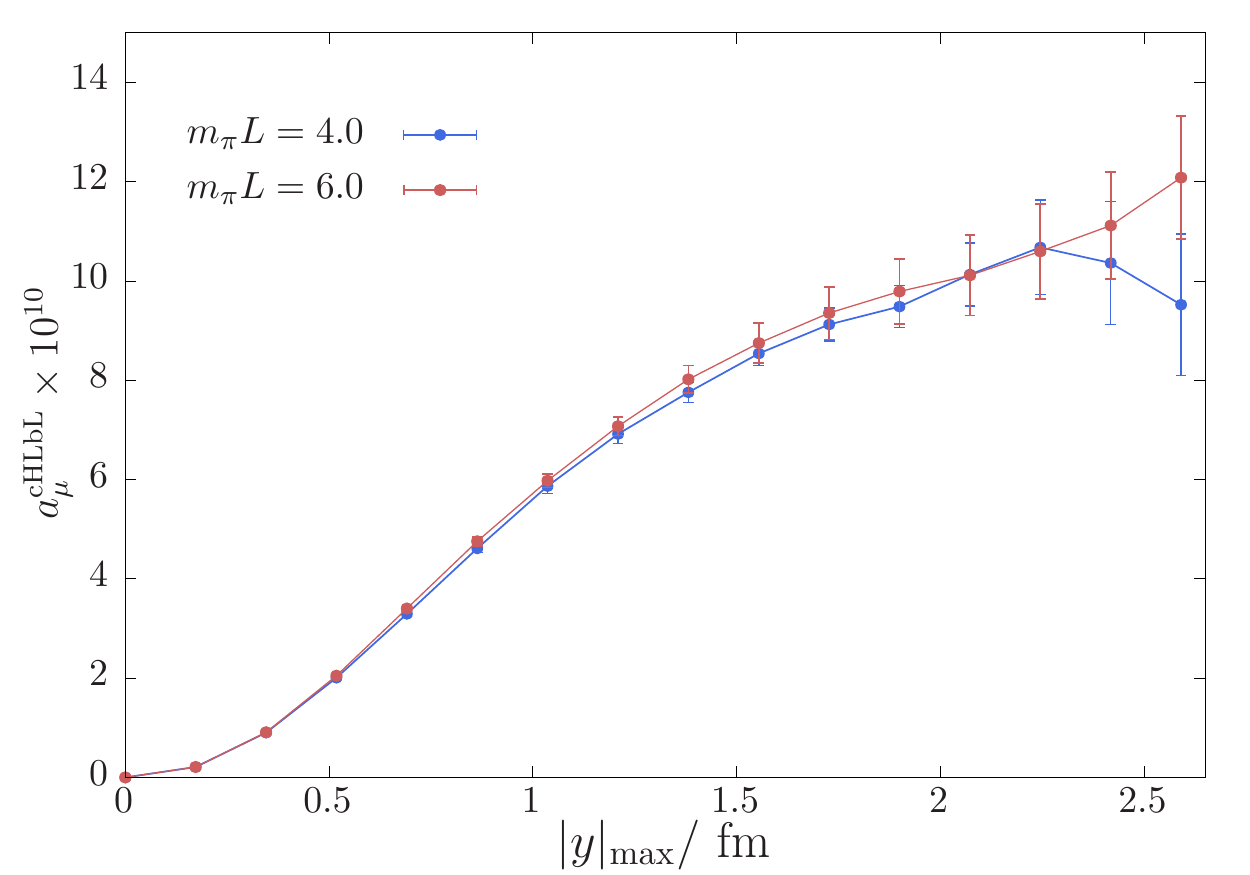}}
   \\
   \subfloat[\label{fig:massdep}]{\includegraphics[width=0.48\textwidth]{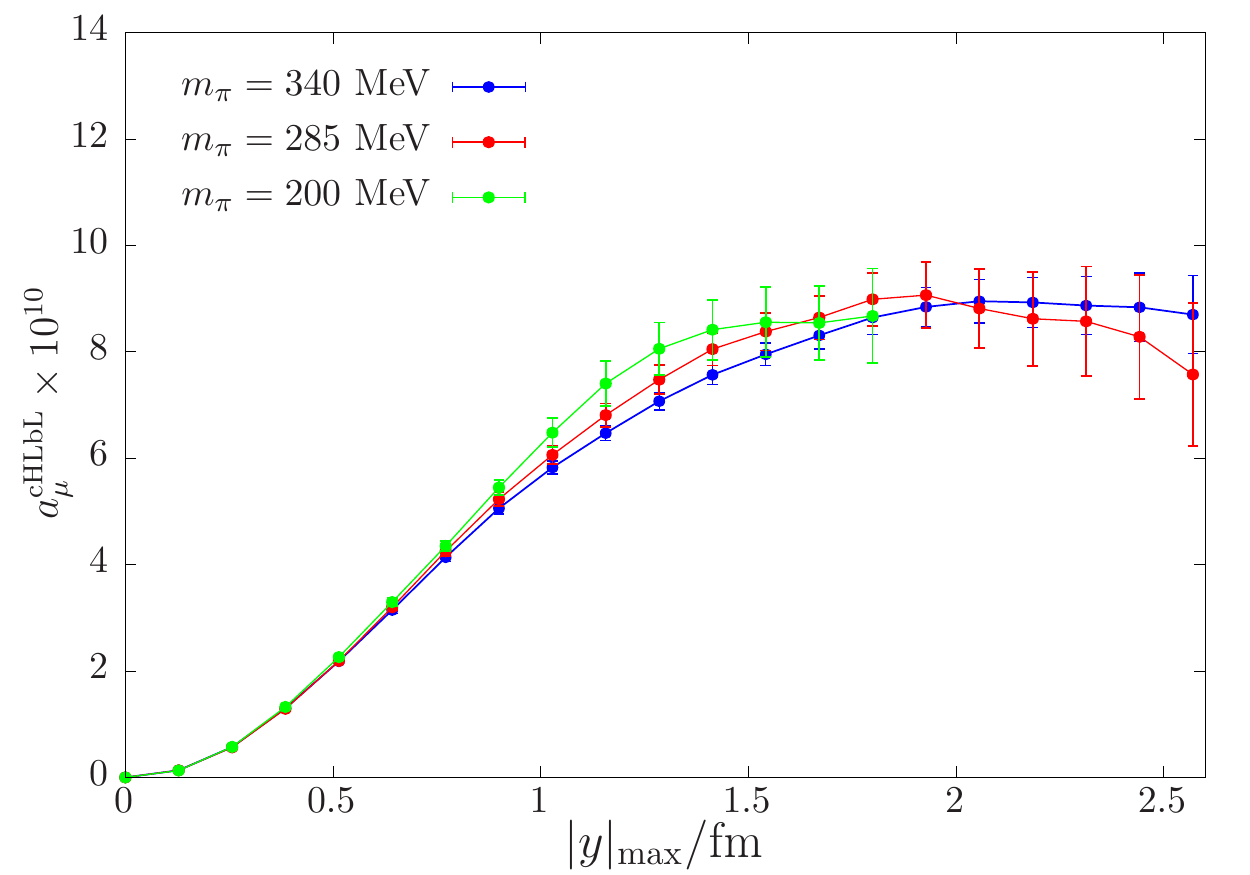}}
\hfill
      \subfloat[\label{fig:contvolfs}]{\includegraphics[width=0.48\textwidth]{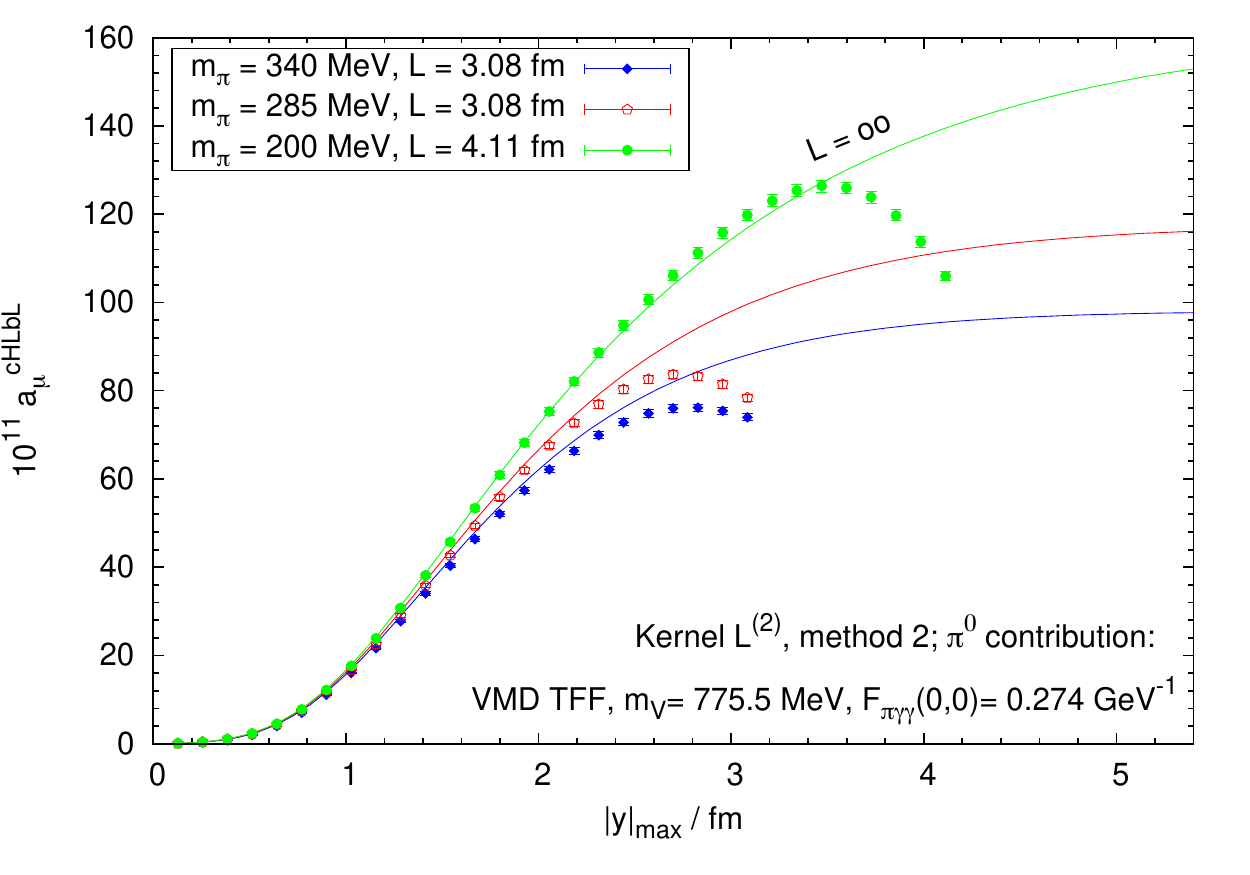}}
   \caption{%
      (a) Discretization effects on the lattices H102 and N203.
      (b) Finite-size effects on the H105 and N101 lattices.
      (c) Pion-mass dependence on the N203, N200 and D200 lattices.
      (d) Pion-mass dependence and finite-size effects for the $\pi^0$-pole contribution.
      The curves represent infinite-volume and the points finite-volume results.
   }
\end{figure}

The ensembles N203 and H102 have a similar pion mass of about \SI{350}{MeV} but 
differ by their lattice spacing and the physical volume of the boxes. At this 
pion mass, the discretization effects can be resolved, 
cf.~\cref{fig:discretization}.  As the ensembles N203, N200, and D200 all have 
the same lattice spacing, comparing them allows us to explore the pion-mass 
dependence of $a_\mu$, which exhibits a mild increase with decreasing pion mass, 
see~\cref{fig:massdep}.  Finite-volume effects become more relevant at long 
distances and precise knowledge of the long-distance tail is very important.  As 
H105 and N101 differ only in their physical volume, a comparison of their 
long-distance behavior allows us to understand the magnitude of our 
finite-volume effects.  These two ensembles seem roughly consistent at large 
distances in~\cref{fig:finitesize}, although their error bars indicate that more 
statistics are~needed.

\begin{table}[h!]
   \centerline{%
      \begin{tabular}{lclcccr}
         \toprule
         Label & $L^3\times T$   & $a~$[fm] & $m_{\pi}~$[MeV] & $m_{\pi}L$ & $L$~[fm] & $\#$confs \\
         \midrule
         H102  & $32^3\times96$  & 0.08636  & $354(5)$        & 5.0        & 2.8      &  900      \\  
         H105  & $32^3\times96$  &          & $284(4)$        & 3.9        & 2.8      & 1000      \\  
         N101  & $48^3\times128$ &          & $282(4)$        & 5.9        & 4.1      &  400      \\
         \midrule
         N203  & $48^3\times128$ & 0.06426  & $345(4)$        & 5.4        & 3.1      &  750      \\  
         N200  & $48^3\times128$ &          & $282(3)$        & 4.4        & 3.1      &  800      \\ 
         D200  & $64^3\times128$ &          & $200(2)$        & 4.2        & 4.1      & 1100      \\ 
         \bottomrule
       \end{tabular}%
    }
    \caption{CLS \(N_f=2+1\) ensembles used in this work.}
    \label{tab:ensembles}
 \end{table}

\section{Pion mass dependence and finite-size effects}
The lattice results presented in the previous section exhibit a mild upward 
trend for decreasing pion mass.  For the neutral pion pole prediction calculated 
in finite volume we obtain a similar behavior. However in infinite volume the 
integral extends to longer distances and correspondingly shows a stronger 
increase as $m_\pi$ is reduced;
see~\cref{fig:contvolfs}. This illustrates the importance of understanding the tail of the integrand
semi-analytically.

We have thus identified two sources of finite volume effects: one is
the truncation of the \(y\)-integral, and the other comes from the
finite-size effect on the lattice integrand itself.  Both artifacts
can be corrected for by semi-analytic continuum computations  (\cref{fig:contvolfs}).
In the small-distance regime the corrections are small and
the lattice data can be used directly. For longer distances, where the
finite-size effect becomes larger, the
$\pi^0$-pole contribution becomes increasingly dominant and we can use the
continuum computation to model the long-distance tail of the \(y\)
integrand.

\begin{figure}[t]
   \begin{center}
      \includegraphics[width=0.48\textwidth,
      trim=0 0 0 1.9em, clip]{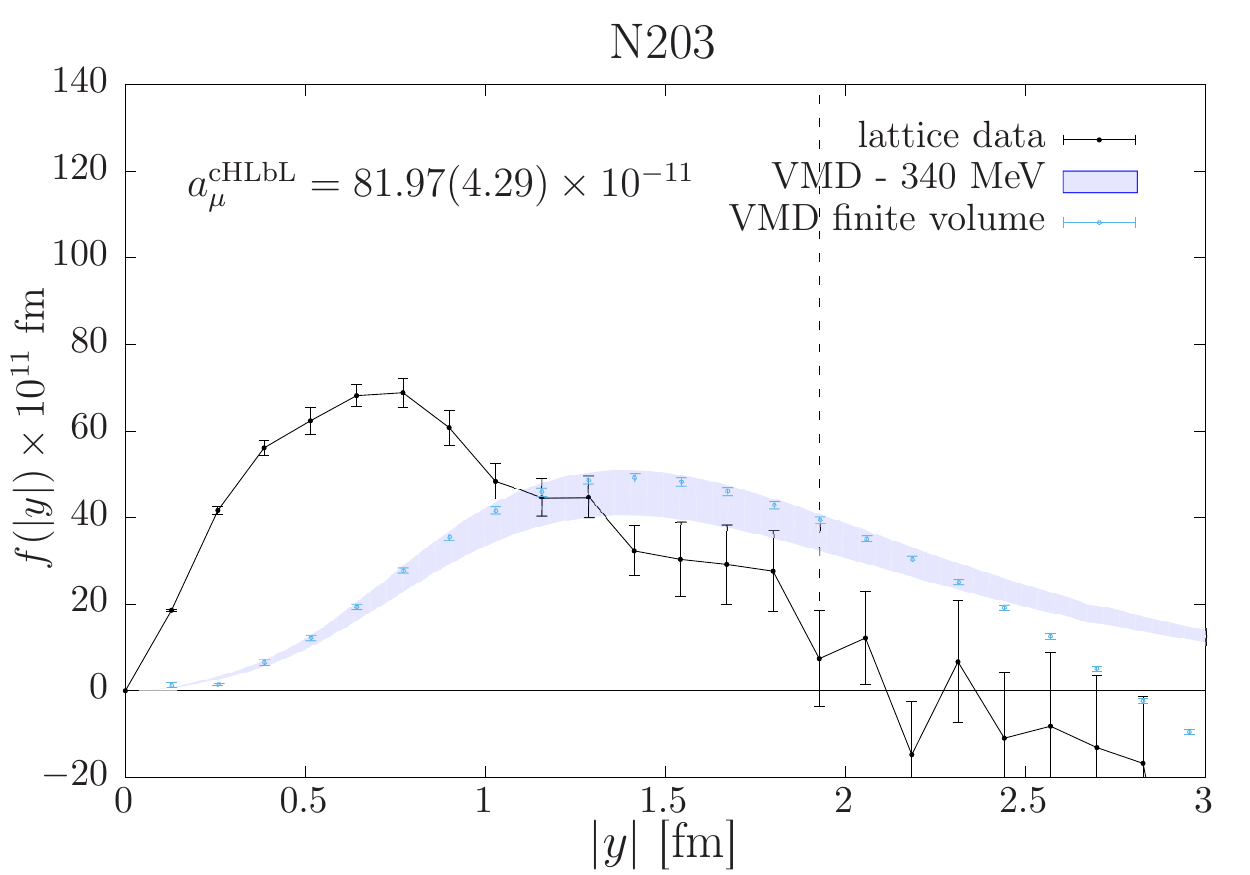}\hfill
      \includegraphics[width=0.48\textwidth]{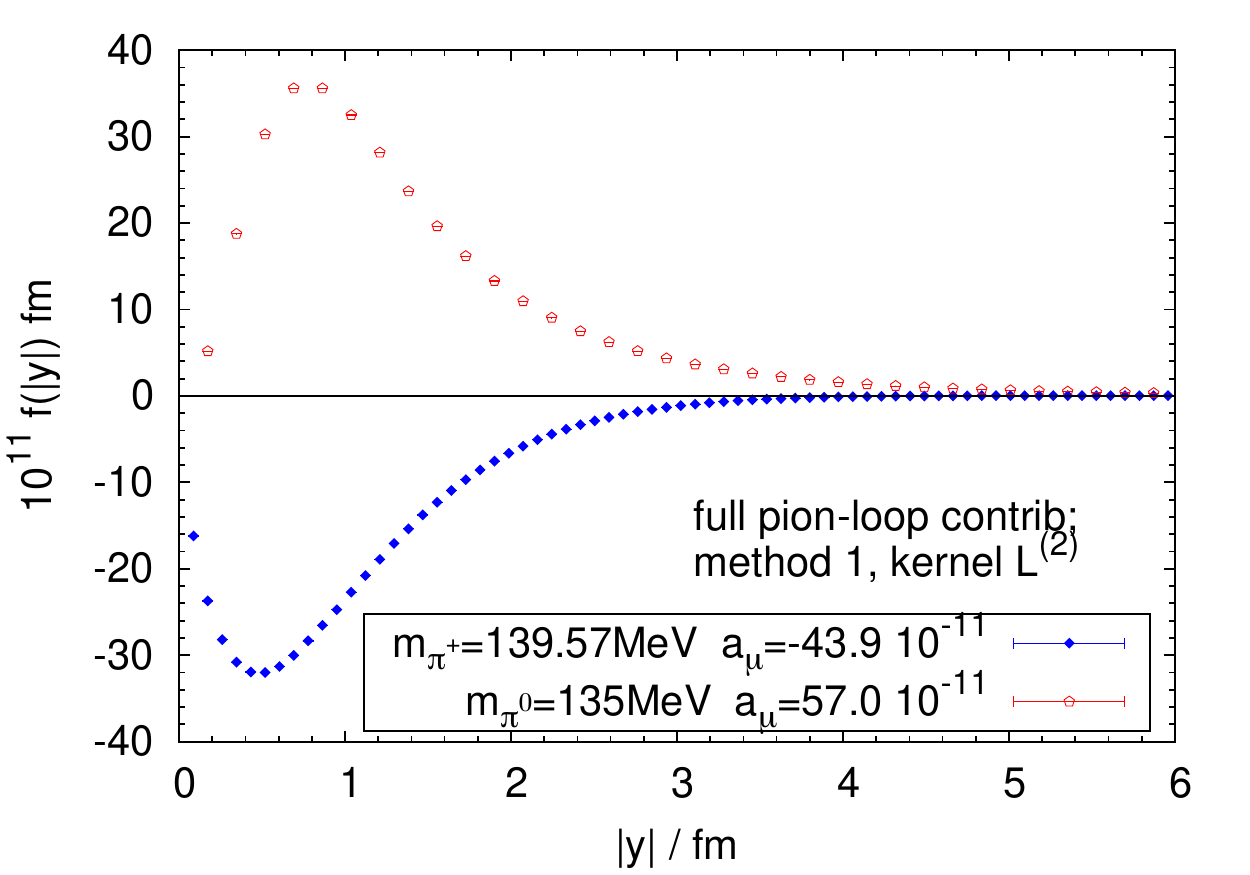}
   \end{center}
   \caption{Left: comparison between the integrand from ensemble N203 and the 
      integrand for the $\pi^0$-pole contribution, both computed with method 2
 and kernel ${\cal L}^{(2)}$. The infinite-volume band covers the normalization factors from $3$ to $34/9$~\cite{Bijnens:2016hgx}, 
 the $\pi^0$ finite-volume points are normalized with the factor $34/9$.
      Right:  pion-pole contribution with a VMD transition form factor (red) and charged 
      pion loop contribution in scalar QED (blue) to $a_\mu^{\rm HLbL}$, using method 1.}
   \label{fig:latvmd}
\end{figure}

\Cref{fig:latvmd} shows the lattice integrand for the N203 lattice and
the corresponding integrand for the $\pi^0$-pole contribution, also computed 
with method 2.  We note that $\Pi^{(1)}_{\mu\nu\sigma\lambda}(x,y,z)$ does not 
contain the pion-level diagram in which the $\pi^0$ propagates between the pair 
$(0,y)$ and the pair $(x,z)$ of vertices, and that the normalization of the two 
other $\pi^0$-pole diagrams is such that
$\tilde\Pi^c_{\mu\nu\sigma\lambda}$ contains the same $\pi^0$ contribution as
$\tilde\Pi_{\mu\nu\sigma\lambda}$, enhanced (in the SU(2)$_{\rm f}$ case) by the charge factor $34/9$.
In \cref{fig:latvmd} we observe effects that are not described by the 
\(\pi^0\)-pole prediction at short distances. At larger distances, we need to 
collect more statistics to test against the $\pi^0$-pole prediction.  The data 
lie below the prediction, suggesting
that there may be a negative contribution to the integrand that is
non-negligible at $|y|=1.5{\rm \,fm}$.

One contribution  to $a_\mu^{\rm HLbL}$ that is known to be negative 
is the charged pion loop. It is also parametrically leading in the chiral limit.
Starting from scalar QED in Euclidean space,
\begin{align}
   \mathcal L_{\rm E}=(\partial_\mu+ieA_\mu)\phi^*(\partial_\mu-ieA_\mu)\phi + m^2 \phi^*\phi
   +\frac{1}{4}F_{\mu\nu}F_{\mu\nu},
\end{align}
we have performed such a computation in our position-space formulation and successfully reproduced the
known charged pion loop contribution~\cite{Kuhn:2003pu}.  The integrand corresponding to
method 1 is shown in the right panel of ~\cref{fig:latvmd} for physical pion masses. Indeed the pion loop
contribution is of comparable size and opposite in sign to the neutral pion pole 
contribution. It is also of shorter range, and if it were further suppressed by 
realistic form factors, it would be unlikely to produce the negative 
contribution suggested by the left panel of \cref{fig:latvmd}.

\section{Conclusions}
The Mainz position-space approach is a method for computing
$a_\mu^{\text{HLbL}}$ using continuum, infinite-volume QED combined
with lattice QCD.  The correctness of the kernel has by now
successfully been tested on the fermion loop, the pion loop as well as
on the neutral pion pole contribution.  Semi-analytic computations based on the
$\pi^0$-pole contribution in finite volume are important to control the
artifacts that stem from the finite size of the box.

The freedom one has in choosing the QED kernel without affecting
$a_\mu^{\text{HLbL}}$ allows for a suppression of certain
discretization effects via subtractions; see Eqs.\ (\ref{eq:L2}--\ref{eq:L3}).  However, the
finite-size effects then turn out to be challenging. Therefore we are
investigating the benefit of a new class of kernels
\be
{\cal L}^{(2,\bar\lambda)}_{\rho\sigma;\mu\nu\lambda} = {\cal L}^{(0)}_{\rho\sigma;\mu\nu\lambda}(x,y)
 - \partial_\mu^{(x)}\Big( x_\alpha \,e^{-\bar\lambda m_\mu^2 x^2/2}\Big)\; {\cal L}^{(0)}_{\rho\sigma;\alpha\nu\lambda}(0,y)
 - \partial_\nu^{(y)}\Big( y_\alpha \,e^{-\bar\lambda m_\mu^2 y^2/2}\Big)\; {\cal L}^{(0)}_{\rho\sigma;\mu\alpha\lambda}(x,0),
\ee
which reduces to ${\cal L}^{(2)}$ for $\bar\lambda=0$, shares its property of vanishing whenever $x$ or $y$ does,
but does not qualitatively alter the long-distance behavior of the original 
kernel ${\cal L}^{(0)}$.
\medskip

{
\noindent {\bf Acknowledgments~~}
This work is supported by the Deutsche Forschungsgemeinschaft (DFG) through the Collaborative Research Centre 1044 and the European Research Council (ERC) under the European Union's Horizon 2020 research and innovation programme through grant agreement 771971-SIMDAMA.
}

\end{document}

%% file: Feynman/methods.tex
   \subfloat[]{%
      \begin{tabular}{l}
         \begin{tikzpicture}[yscale=0.10, xscale=0.10]
            \begin{feynman}
               \vertex (x) at ( 00, 00);
               \vertex (y) at ( 05,-05);
               \vertex (w) at ( 10, 00);
               \vertex (z) at ( 05, 05);
               \vertex (ml) at (-05,-10);
               \vertex (mr) at ( 15,-10);
               \vertex (mx) at ( 00,-10);
               \vertex (my) at ( 05,-10);
               \vertex (mw) at ( 10,-10);
               \vertex (mz) at ( 05, 05);
               \vertex (hz) at ( 05, 10);
               \diagram*{
                  (ml) -- (mr),
                  (x) -- (y) -- (z) -- (w) -- (x),
                  (mx) -- [photon] (x),
                  (my) -- [photon] (y),
                  (mw) -- [photon] (w),
                  (mz) -- [photon] (z),
                  (hz) -- [photon] (z),
               };
            \end{feynman}
            \begin{feynman}
               \vertex (x) at ( 25, 00);
               \vertex (y) at ( 30,-05);
               \vertex (w) at ( 35, 00);
               \vertex (z) at ( 30, 05);
               \vertex (ml) at ( 20,-10);
               \vertex (mr) at ( 40,-10);
               \vertex (mx) at ( 25,-10);
               \vertex (my) at ( 30,-10);
               \vertex (mw) at ( 35,-10);
               \vertex (mz) at ( 30, 05);
               \vertex (hz) at ( 30, 10);
               \diagram*{
                  (ml) -- (mr),
                  (y) -- (x) -- (z) -- (w) -- (y),
                  (mx) -- [photon] (x),
                  (my) -- [photon] (y),
                  (mw) -- [photon] (w),
                  (mz) -- [photon] (z),
                  (hz) -- [photon] (z),
               };
            \end{feynman}
            \begin{feynman}
               \vertex (x) at ( 50, 00);
               \vertex (y) at ( 55,-05);
               \vertex (w) at ( 60, 00);
               \vertex (z) at ( 55, 05);
               \vertex (ml) at ( 45,-10);
               \vertex (mr) at ( 65,-10);
               \vertex (mx) at ( 50,-10);
               \vertex (my) at ( 55,-10);
               \vertex (mw) at ( 60,-10);
               \vertex (mz) at ( 55, 05);
               \vertex (hz) at ( 55, 10);
               \diagram*{
                  (ml) -- (mr),
                  (y) -- (z) -- (x) -- (w) -- (y),
                  (mx) -- [photon] (x),
                  (my) -- [photon] (y),
                  (mw) -- [photon] (w),
                  (mz) -- [photon] (z),
                  (hz) -- [photon] (z),
               };
            \end{feynman}
         \end{tikzpicture}
   \\
         \begin{tikzpicture}[yscale=0.10, xscale=0.10]
            \begin{feynman}
               \vertex (x) at ( 00, 00);
               \vertex (y) at ( 05,-05);
               \vertex (w) at ( 10, 00);
               \vertex (z) at ( 05, 05);
               \vertex (ml) at (-05,-10);
               \vertex (mr) at ( 15,-10);
               \vertex (mx) at ( 00,-10);
               \vertex (my) at ( 05,-10);
               \vertex (mw) at ( 10,-10);
               \vertex (mz) at ( 05, 05);
               \vertex (hz) at ( 05, 10);
               \diagram*{
                  (ml) -- (mr),
                  (x) -- (y) -- (z) -- (w) -- (x),
                  (mx) -- [photon] (x),
                  (my) -- [photon] (y),
                  (mw) -- [photon] (w),
                  (mz) -- [photon] (z),
                  (hz) -- [photon] (z),
               };
            \end{feynman}
            \begin{feynman}
               \vertex (x) at ( 25, 00);
               \vertex (y) at ( 30,-05);
               \vertex (w) at ( 35, 00);
               \vertex (z) at ( 30, 05);
               \vertex (ml) at ( 20,-10);
               \vertex (mr) at ( 40,-10);
               \vertex (mx) at ( 25,-10);
               \vertex (my) at ( 30,-10);
               \vertex (mw) at ( 35,-10);
               \vertex (mz) at ( 30, 05);
               \vertex (hz) at ( 30, 10);
               \diagram*{
                  (ml) -- (mr),
                  (x) -- (y) -- (z) -- (w) -- (x),
                  (mx) -- [photon] (y),
                  (my) -- [photon] (x),
                  (mw) -- [photon] (w),
                  (mz) -- [photon] (z),
                  (hz) -- [photon] (z),
               };
            \end{feynman}
            \begin{feynman}
               \vertex (x) at ( 50, 00);
               \vertex (y) at ( 55,-05);
               \vertex (w) at ( 60, 00);
               \vertex (z) at ( 55, 05);
               \vertex (ml) at ( 45,-10);
               \vertex (mr) at ( 65,-10);
               \vertex (mx) at ( 50,-10);
               \vertex (my) at ( 55,-10);
               \vertex (mw) at ( 60,-10);
               \vertex (mz) at ( 55, 05);
               \vertex (hz) at ( 55, 10);
               \diagram*{
                  (ml) -- (mr),
                  (x) -- (y) -- (z) -- (w) -- (x),
                  (mx) -- [photon, bend right] (w),
                  (my) -- [photon] (y),
                  (mw) -- [photon, bend left] (x),
                  (mz) -- [photon] (z),
                  (hz) -- [photon] (z),
               };
            \end{feynman}
         \end{tikzpicture}%
      \end{tabular}%
   }